\documentclass{ws-ijn}
\usepackage{multicol}
\def\wsdraftnote{}
\begin{document}

\catchline{X}{X}{2019}{}{}
\markboth{C.~A.~Downing and M.~E.~Portnoi}{Trapping charge carriers in low-dimensional Dirac semimetals}

\title{Trapping charge carriers in low-dimensional Dirac materials}

\author{C.~A.~Downing}
\address{Departamento de F\'{i}sica de la Materia Condensada,
CSIC-Universidad de Zaragoza,\\
Zaragoza 50009, Spain\\
\email{downing@unizar.es}}

\author{M.~E.~Portnoi}
\address{School of Physics, University of Exeter, Stocker Road, Exeter EX4 4QL, United Kingdom\\
\& ITMO University, St. Petersburg 197101, Russia\\
\email{m.e.portnoi@exeter.ac.uk}}

\maketitle

\begin{history}
\received{19 January 2019}
\revised{XX XXX 2019}
\end{history}

\begin{abstract}

We consider the problem of confining the famously elusive Dirac-like quasiparticles, as found in some recently discovered low-dimensional systems. After briefly surveying the existing theoretical proposals for creating bound states in Dirac materials, we study relativistic excitations with a position-dependent mass term. With the aid of an exactly-solvable model, we show how bound states begin to emerge after a critical condition on the size of the mass term is met. We also reveal some exotic properties of the unusual confinement discovered, including an elegant chevron structure of the bound state energies as a function of the size of the mass.

\end{abstract}

\keywords{Dirac materials, bound states, Dirac equation, confinement, relativistic wave equation}

\begin{multicols}{2}

\section{On the problem of charge confinement}
\label{sec:intro}

Since the discovery of graphene in 2004, the exciting field of Dirac materials (those which host excitations who obey a version of Dirac's relativistic equation for the electron) was born\cite{Thaller1956} \cite{Shen2012} \cite{Bagrov2014}. These exotic materials, where relativistic quantum mechanics and condensed matter physics merge in a nontrivial way, are a playground for groundbreaking fundamental discoveries, as well as for the next generation of electronic devices\cite{Wehling2014}.

A key issue to address when considering the electronic properties of Dirac materials is how to confine their relativistic charge carriers. This is because these quasi-particles exhibit Klein (chiral) tunneling effects\cite{Klein1929}, leading to high transmission probabilities in standard potential step barrier problems\cite{Katsnelson2006}. For the case of massless (ultrarelativistic) particles, the situation seems even more desperate, since their governing wave equation in potential wells maps onto the Schr\"{o}dinger equation for scattering states, suggesting that the generation of bound states is an impossible task\cite{Tudorovskiy2007}.

In this work, we briefly review some of the theoretical suggestions for creating discrete states in Dirac materials. We then propose a theory of bound states of Dirac particles with a position-dependent mass term. Guided by an exactly-solvable model, we reveal how bound states first appear (at zero energy) at a threshold strength of the mass term; and how further bound states appear regularly in a remarkable chevron pattern. We also characterize, with a brief equation, the number of bound states in the system as a function of the size of the mass. Therefore, our work opens up another avenue to explore in the quest to trap Dirac-like charge carriers.

In what follows, we review some theoretical proposals which enable bound states of Dirac particles in Sec.~\ref{sec:literature}, and we introduce our theory of bound states of relativistic particles with a position-dependent mass in Sec.~\ref{sec:bound}, before drawing some brief conclusions in Sec.~\ref{sec:conc}.

\section{Some previous proposals to bind relativistic particles}
\label{sec:literature}

Here we briefly outline some theoretical proposals for achieving bound states in Dirac materials\cite{Rozhkov2011} \cite{Review2019}. We focus on four principle themes: trapping in magnetic fields in Sec.~\ref{sec:mag}; confinement of zero energy states in electric fields in Sec.~\ref{sec:zero}; bound states due to precise Fermi velocity engineering in Sec.~\ref{sec:fermi}; and finally we discuss the crucial role of the mass term in Sec.~\ref{sec:mass}.

\subsection{Magnetic fields}
\label{sec:mag}

Perhaps the simplest method to bring about the on-demand confinement of Dirac particles is through the application of magnetic fields perpendicular to the Dirac material. Of course, a constant magnetic field profile leads to the emergence of relativistic Landau levels and the quantum Hall effect\cite{Zhang2005}. So-called magnetic quantum dots,\cite{Martino2007} where the magnetic field is spatially inhomogeneous, have also been considered recently and shown to lead to the formation of bound states,\cite{Ramezani2009} \cite{Roy2012} albeit under some quite strict criteria for the decay strength of the magnetic field\cite{Downing2016} \cite{Downing2016b}. The experimental realization of magnetic quantum dots in Dirac materials is an open task.

\subsection{Electric fields}
\label{sec:zero}

Confinement in quantum dots formed by potential wells has been extensively studied\cite{Silvestrov2007} \cite{Matulis2008} \cite{Recher2010}. For massive particles, bound states are allowed despite the Klein phenomenon, as has been shown by a plethora of exact solutions of Dirac-like equations in various potential wells \cite{Downing2014} \cite{Hartmann2014} \cite{Hartmann2017} \cite{Hartmann2017b}. However, the situation is very different for massless particles\cite{Hartmann2010} \cite{Downing2011}. Then the Klein tunneling always succeeds and only quasi-bound states may form. However, a series of works has shown truly bound states may form at zero-energy, both in two-dimensional\cite{Bardarson2009} \cite{Downing2015} \cite{Portnoi2017} and one-dimensional\cite{Hartmann2011} \cite{Stone2012} Dirac materials. These works exploit the fact that the pseudospin, the conserved quantity which allows for perfect chiral transmission, is ill-defined at the apex of the Dirac cone, that is, precisely at zero energy. The experimental signatures of zero energy bound states have been recently reported in groundbreaking measurements\cite{Lee2016} \cite{Bai2018}.

\subsection{Fermi velocity engineering}
\label{sec:fermi}

It has been put forward that Dirac particles may also be confined by so-called velocity barriers\cite{Peres2009}. This term refers to barriers formed by a region of one Fermi velocity surrounded by a region of a different Fermi velocity, leading to an exotic form of trapping potential. A succession of theoretical papers have described the scattering and bound state solutions in such circumstances\cite{Concha2010} \cite{Raoux2010} \cite{Downing2017b}, which may arise in certain strain configurations of the Dirac material, as shown in the latest experimental work \cite{Downs2016}.

\subsection{Inducing a mass term}
\label{sec:mass}

It has already been mentioned in Sec.~\ref{sec:zero} that confinement in quantum dots is much easier when the Dirac particles are massive, that is, when the Hamiltonian contains a term proportional to the third Pauli matrix $\sigma_z$. Alternatively, if one is dealing with massless particles, one may employ the so-called Berry-Mondragon (or infinite-mass) boundary condition\cite{Berry1987}. This condition may confine ultrarelativistic excitations to a specific small region, outside of which their mass is considered infinite. Such a condition has wide applications in the latest nanoribbon and nanotube research, where edge termination is of particular importance\cite{Saroka2014} \cite{Saroka2015} \cite{Saroka2016}.

\section{Bound states of Dirac particles with a position-dependent mass}
\label{sec:bound}

Here we consider Dirac particles in one-dimension with a spatially inhomogeneous mass term, in a theoretical framework outlined in Sec.~\ref{sec:ham}. Bound states of quantum particles with a position dependent mass have been studied before, predominately in the context of non relativistic systems\cite{Dekar1998} \cite{Alhaidari2002} \cite{Koc2003}, but also in the relativistic regime\cite{Alhaidari2004}. Here we show some remarkable properties of bound states of relativistic particles with a position-dependent mass, with the aid of an elegant exactly-solvable model introduced in Sec.~\ref{sec:model}.

\subsection{The Hamiltonian}
\label{sec:ham}

The Hamiltonian of a massive relativistic particle in a one-dimensional Dirac material reads
\begin{equation}
\label{eq1}
H = v_{\mathrm{F}} p_x \sigma_x + m  v_{\mathrm{F}}^2 \sigma_z,
\end{equation}
where the Fermi velocity $v_{\mathrm{F}}$ plays the role of the speed of light $c$ from Dirac's theory, $p_x$ is the momentum operator and $m$ is the quasiparticle mass. In Eq.~\eqref{eq1}, $\sigma_{x, z}$ are Pauli's first and third spin matrices respectively. The Hamiltonian~\eqref{eq1} describes Dirac particles with the famous relativistic spectrum
\begin{equation}
\label{eq2}
E = \pm \sqrt{ \left( m v_{\mathrm{F}}^2 \right)^2 +  \left( v_{\mathrm{F}} p_x \right)^2},
\end{equation}
where $\pm$ alludes to electron and hole-like quasiparticles, which mimics the electron and positron symmetry from Dirac's original theory of the relativistic electron. Let us now consider a position dependent mass,
\begin{equation}
\label{eq2b}
m = m(x),
\end{equation}
in the $2 \times 2$ matrix Hamiltonian~\eqref{eq1}, which leads to the Schr\"{o}dinger equation
\begin{equation}
\label{eq3}
H \Psi = E \Psi, \quad \Psi (x) = \left(
 \begin{array}{c}
\psi_A(x) \\ \psi_B(x)
 \end{array}
\right),
\end{equation}
where the wavefunction $\Psi$ is a two-component column vector, reflecting the pseudospin of the system. Therefore, one is presented with the following system of two coupled ordinary differential equations to be solved
\begin{subequations}
\label{eq4}
 \begin{align}
  \frac{d}{dx} \psi_B &= \left( \varepsilon - \Delta (x) \right) \psi_A, \label{eq4a} \\
  \frac{d}{dx} \psi_A &= \left( \varepsilon + \Delta (x) \right) \psi_B, \label{eq4b}
 \end{align}
\end{subequations}
where the reduced energy is $\varepsilon = E / \hbar v_{\mathrm{F}}$ and the reduced mass profile function is $\Delta (x) = m v_{\mathrm{F}} / \hbar$. Switching the  dependent variables with the relations $\psi_{1, 2} = \psi_A \pm \psi_B$, leads to the new system of equations
\begin{subequations}
\label{eq5}
 \begin{align}
  \left( \varepsilon - \frac{d}{dx} \right) \psi_1 &=  \Delta (x) \psi_2, \label{eq5a} \\
  \left( \varepsilon + \frac{d}{dx} \right)  \psi_2 &=  \Delta (x) \psi_1, \label{eq5b}
 \end{align}
\end{subequations}
which are in a convenient form for further calculations. Indeed, it is straightforward to show that Eq.~\eqref{eq5} admit analytic solutions for a variety of band-gaps profiles $\Delta (x)$, including functional forms known from other studies\cite{DowningHeun} \cite{Downing2017}, and we will revisit such solutions in a future work. From now on, we label the total spinor wavefunction as $\phi = \left(\psi_1, \psi_2 \right)$ and consider a specific $\Delta (x)$.

\subsection{The model}
\label{sec:model}

We consider a mass term changing exponentially in space, modeled with the symmetric reduced band-gap function 
\begin{equation}
\label{eq6}
\Delta (x) = \Delta_0 e^{-\tfrac{|x|}{d}},
\end{equation}
where $\Delta_0$ is the maximal strength of the band-gap and $d$ is the length scale over which the band-gap decays.  

We begin by considering the system of equations~Eq.~\eqref{eq5} with the mass-profile~\eqref{eq6} in the half-space $x>0$. Upon eliminating $\psi_2$, the second-order differential equation governing $\psi_1$ is given by
\begin{equation}
\label{eq7}
\psi_1 '' (x) + \left( \tfrac{1}{d} \right) \psi_1 ' (x) + \left( \Delta_0 e^{-\tfrac{2 x}{d}} - \varepsilon^2 - \tfrac{\varepsilon}{d} \right) \psi_1 (x) = 0,
\end{equation}
where the prime denotes taking a derivative with respect to $x$. Changing the independent variable to
\begin{equation}
\label{eq7b}
z = e^{-\tfrac{x}{d}},
\end{equation}
leads to the transformation of Eq.~\eqref{eq7} into
\begin{equation}
\label{eq8}
\psi_1 '' (z) + \left( \left[ \Delta_0 d \right]^2 - \tfrac{\varepsilon d \left[ 1 + \varepsilon d \right]}{z^2} \right) \psi_1 (z) = 0, 
\end{equation}
where the prime now refers to the new variable $z$, and where the term proportional to $\psi_1 '$ has been eliminated. In Eq.~\eqref{eq8}, the ansatz
\begin{equation}
\label{eq9}
\psi_1 (z) = \sqrt{z} f(z), 
\end{equation}
gives rise to the following equation for the unknown function $f(z)$:
\begin{equation}
\label{eq10}
f '' (z) + \left( \tfrac{1}{z} \right) f' (z) + \left( \left[ \Delta_0 d \right]^2 - \tfrac{ \left[ \varepsilon d + 1/2 \right]^2}{z^2} \right) f (z) = 0.
\end{equation}
This equation is simply Bessel's differential equation in the scaled variable $\Delta_0 d \: z$, and thus has the standard solution
\begin{equation}
\label{eq10b}
f(z) = c_1 J_{\varepsilon d + 1/2} \left( \Delta_0 d \: z \right) + c_2 Y_{\varepsilon d + 1/2} \left( \Delta_0 d \: z \right),
\end{equation}
where $J_{\alpha} \left( \xi \right)$ and $Y_{\alpha} \left( \xi \right)$ are the Bessel functions of the first and second kinds respectively, both of order $\alpha$, and $c_{1, 2}$ are arbitrary constants. We consider positive energies $\varepsilon >0$ for the moment, and always require $\psi_1$ to be square-integrable, so that the physical solution to Eq.~\eqref{eq10} is with $J_{\alpha} \left( \xi \right)$ only. Along with Eq.~\eqref{eq7b} and Eq.~\eqref{eq9}, this selection from Eq.~\eqref{eq10b} completes the search for $\psi_1(x)$ for $x > 0$.

Recalling the coupled equation~\eqref{eq5}, once $\psi_1$ is known, $\psi_2$ directly follows from Eq.~\eqref{eq5a}. Hence the complete spinor wavefunction in the half-space $x>0$ reads
\begin{equation}
\label{eq11}
 \phi^{x > 0} (x) = \frac{C}{\sqrt{d}}  e^{-\tfrac{x}{2 d}} \left(
 \begin{array}{c}
 J_{\varepsilon d + 1/2} \left( \Delta_0 d \: e^{-\tfrac{x}{d}} \right) \\
 J_{\varepsilon d - 1/2} \left( \Delta_0 d \: e^{-\tfrac{x}{d}} \right)
 \end{array}
\right),
\end{equation}
where $C$ is a constant to be determined by normalization of the total wavefunction $\phi(x)$ over all space. The symmetries of the coupled equations~\eqref{eq5} allow one to then write down the corresponding solution in the half-space $x<0$ as
\begin{equation}
\label{eq12}
 \phi^{x < 0} (x) = \tau \frac{C}{\sqrt{d}}  e^{\tfrac{x}{2 d}} \left(
 \begin{array}{c}
 J_{\varepsilon d - 1/2} \left( \Delta_0 d \: e^{\tfrac{x}{d}} \right) \\
 J_{\varepsilon d + 1/2} \left( \Delta_0 d \: e^{\tfrac{x}{d}} \right)
 \end{array}
\right),
\end{equation}
where we have introduced the parameter
\begin{equation}
\label{eq12b}
 \tau = \pm 1.
\end{equation}
After imposing the continuity of both wavefunction components $\psi_1$ and $\psi_2$ at the interface $x=0$, the following condition arises
\begin{equation}
\label{eq13}
 J_{\varepsilon d - 1/2} \left( \Delta_0 d \right) = \tau J_{\varepsilon d + 1/2} \left( \Delta_0 d \right),
\end{equation}
which is the exact expression governing the allowed eigenenergies $\varepsilon$ of the system, as a function of the two mass profile parameters $\Delta_0$ and $d$ [c.f. Eq.~\eqref{eq6}]. In the limit of small energy $\varepsilon d \ll 1$, Eq.~\eqref{eq13} simplifies considerably into the analytic expression
\begin{equation}
\label{eq14}
  \Delta_0 d = \pi \left( n_{\tau} + \tfrac{\tau}{4} \right), \quad \varepsilon d \ll 1,
\end{equation}
where $n_{+} = 0, 1, 2...$ and $n_{-} = 1, 2, 3...$ are integers and $\tau$ is given by Eq.~\eqref{eq12b}. Equation~\eqref{eq14} simply governs the existence of zero energy ($\varepsilon = 0$) bound states at critical band-gap sizes. Away from this limit, one may find the exact bound state energies from the roots of Eq.~\eqref{eq13}. We note a similar equation to Eq.~\eqref{eq13} can be derived when considering negative energies $\varepsilon <0$.

The result of standard root-finding procedures for the eigenvalue equation~\eqref{eq13} is shown in Fig.~\ref{fig1}, which displays a beautiful chevron structure. It is immediately striking that bound states only occur above a threshold band-gap size, given by
\begin{equation}
\label{eq15}
  \left( \Delta_0 d \right)_{\mathrm{threshold}} = \frac{\pi}{4},
\end{equation} 
which is indeed a zero energy state, described by $n_{+} = 0$ in Eq.~\eqref{eq14}. In Fig.~\ref{fig1}, the dimensionless eigenenergies $\varepsilon d$ are found from Eq.~\eqref{eq13} with $\tau = +$ (red lines) and $\tau = -$ (blue lines), as a function of the dimensionless band-gap strength $\Delta_0 d$ (scaled by $4/\pi \simeq 1.27$ for convenience). All of the bound states fall within the region demarcated by $\varepsilon = \pm \Delta_0$ (dashed black lines). The emergence of new zero energy states at constant intervals of $\Delta_0 d = \pi/2$ [c.f. Eq.~\eqref{eq14}], leads to a neat classification of the total number of bound states $N$ for a system with some finite $\Delta_0 d$. The categorization reads
\begin{subequations}
\label{eq16}
 \begin{align}
  N = 0, \quad &\Delta_0 d < \tfrac{\pi}{4}, \label{eq16a} \\
  N = 1, 3, 5, ..., \quad &\Delta_0 d = \tfrac{N\pi}{4}, \label{eq16b} \\
  N = 2, 4, 6, ..., \quad & \tfrac{\left( N - 1 \right)\pi}{4} < \Delta_0 d < \tfrac{\left( N + 1 \right)\pi}{4}, \label{eq16c} 
 \end{align}
\end{subequations}
such that are no bound states ($N=0$) below a threshold band-gap strength [c.f. Eq.~\eqref{eq15}]; there are an odd number of bound states at critical band-gap strengths [c.f. Eq.~\eqref{eq14}]; and otherwise there are regions of an even number of bound states. This scheme is marked in Fig.~\ref{fig1} by the critical band-gap sizes $\Delta_0 d = \{ \pi/4, 3\pi/4, 5\pi/4, 7\pi/4, 9\pi/4 \}$ (dotted gray lines).

\begin{figurehere}
\centerline{\includegraphics[width=8.0cm]{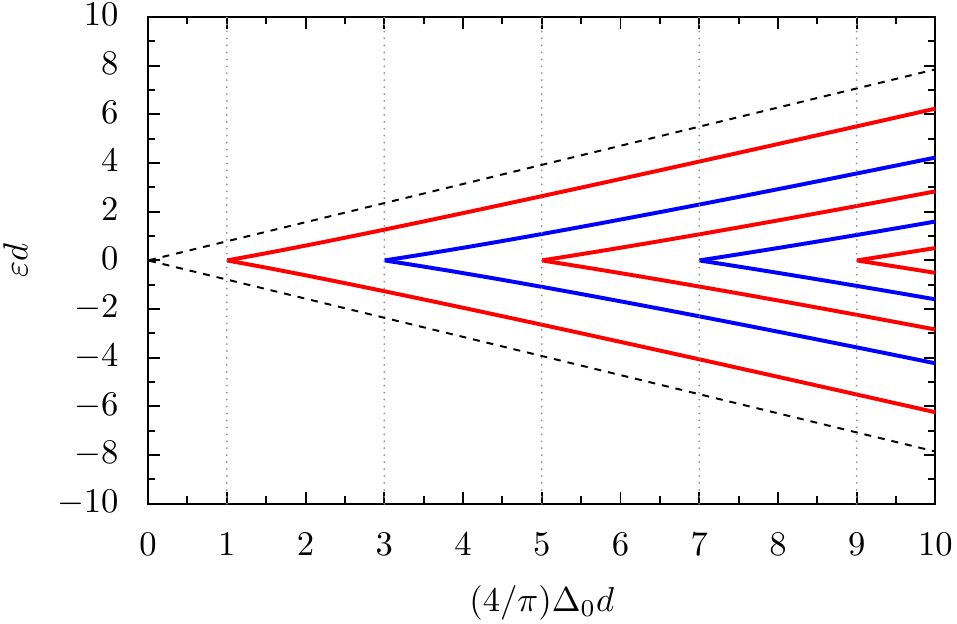}}
\caption{A plot of eigenenergy $\varepsilon d$ against bandgap strength $\Delta_0 d$, as found from Eq.~\eqref{eq13} for $\varepsilon >0$, with $\tau = +$ (red lines) and $\tau = -$ (blue lines). Guides for the eye: $\varepsilon = \pm \Delta_0$ (dashed black lines). Critical bandgaps: $\Delta_0 d = \{ \pi/4, 3\pi/4, 5\pi/4, 7\pi/4, 9\pi/4 \}$ (dotted gray lines).}
\label{fig1}
\end{figurehere}

The results of this section highlight an under-appreciated approach to bind relativistic Dirac particles: confinement by considering a position-dependent mass profile. We have seen some remarkable properties of bound states in such systems, via an exactly-solvable model, including the importance of a threshold strength of band-gap [c.f. Eq.~\eqref{eq14}], and a beautiful chevron structure of the bound states trapped in the effective mass barrier [c.f. Fig.~\ref{fig1}].

\section{Conclusions}
\label{sec:conc}

We have considered the most conceptually simple problem associated with Dirac-like charge carriers in the emerging field of Dirac materials: how can one  confine and manipulate such elusive particles? We have briefly reviewed some theoretical literature on the topic of charge confinement low-dimensional Dirac materials, including the use of magnetic and electric fields, Fermi velocity engineering and the introduction of a mass term.

We have shown, through an analytic model, that Dirac particles with a position-dependent mass profile are able to be confined in certain situations. We have detailed the properties of such bound states, including how they form a remarkable chevron structure. Our work paves the way for more sophisticated theoretical studies of low-dimensional Dirac systems which host excitations with spatially inhomogeneous mass profiles, with the eventual aim of experimental detection of the novel bound states predicted in this work.

\nonumsection{Acknowledgments}
\noindent C.A.D. acknowledges support from the Juan de la Cierva program (MINECO, Spain). M.E.P. is grateful for the support of the EU H2020 RISE project CoExAN (Grant No. H2020-644076). This work was supported by the Government of the Russian Federation through the ITMO Fellowship and Professorship Program.

\end{multicols}
\end{document}